\documentclass{vgtc}                          

\graphicspath{{figures/}{pictures/}{images/}{./}} 

\usepackage{times}                     

\usepackage{tabu}                      
\usepackage{booktabs}                  
\usepackage{lipsum}                    
\usepackage{mwe}                       
\usepackage[normalem]{ulem}

\usepackage{mathptmx}                  
\usepackage{enumitem}
\usepackage{amsmath}

\newcommand{\add}[1]{\textcolor{red}{#1}}
\newcommand{\remove}[1]{\textcolor{red}{\sout{#1}}}
\renewcommand{\add}[1]{#1}
\renewcommand{\remove}[1]{}

\onlineid{1021}

\vgtccategory{Research}

\vgtcinsertpkg

\title{May (A)I Beautify Your Visualization?\\Expert Judgments of Acceptable Aesthetic Alterations}

\author{Kalina Borkiewicz\thanks{e-mail: kalina@sci.utah.edu}
        \\\scriptsize University of Utah %
\and Jixian Li
     \\\scriptsize University of Utah %
\and Joshua A. Levine
     \\\scriptsize University of Arizona %
\and Katherine E. Isaacs
     \\\scriptsize University of Utah}

\teaser{
  \centering
  \includegraphics[width=\linewidth]{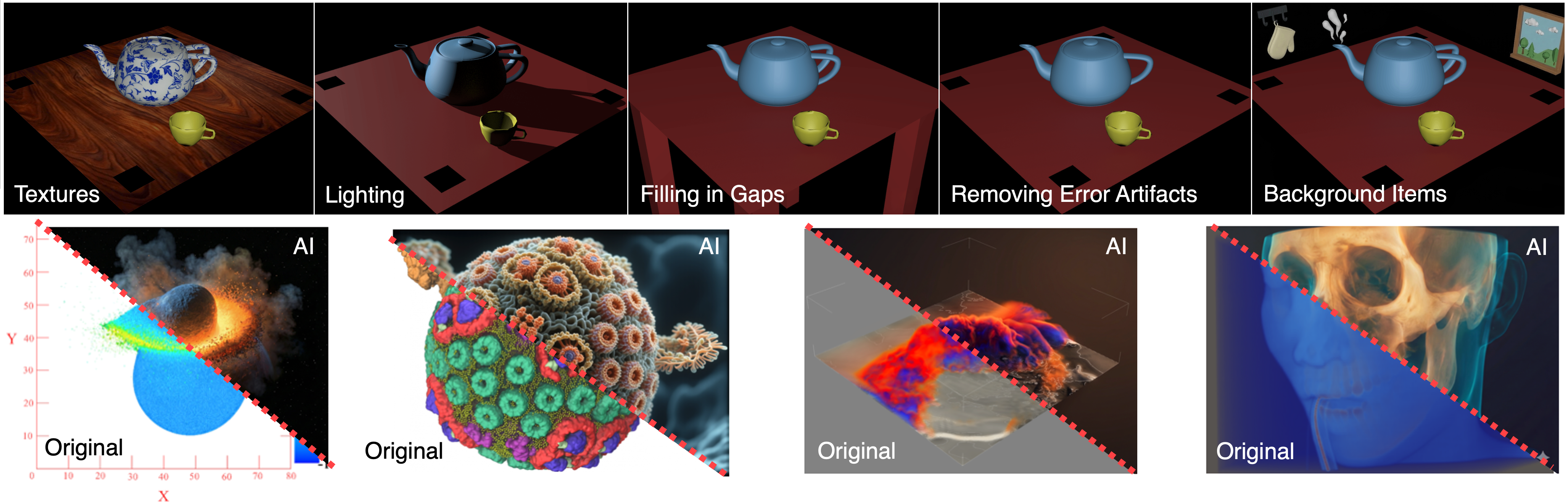}
  \caption{Examples of aesthetic alterations from our study. Top row: a subset of abstract transformation categories. Bottom row: pairs of original (lower-left) and AI-generated (upper-right) visualizations. 
  These examples illustrate a range of alterations that may preserve, enhance, or distort the message or features of the data. \small{\textit{AI images generated with Nano Banana Pro and Qwen-Image.}}}

  \label{fig:teaser}
}

\abstract{
In 3D visualizations of natural phenomena, improving aesthetics can provide measurable benefits, but often involves transformations that affect how the data is perceived. As a growing range of tools---including AI-based methods---make visual design and modification more accessible, it is increasingly important to understand trade offs and concerns when making these changes. We conducted an expert survey (N=95) with visualization researchers, practitioners, and domain scientists, investigating reactions to fifteen alterations spanning presentation-level adjustments (e.g.,~lighting, camera position) and data-level modifications (e.g.,~removing errors, filling gaps), applied by both humans and AI systems. Results show differences in perceived acceptability are driven by the transformation's meaning, regardless of whether it operates at the presentation or data level. Additionally, certain modifications were consistently judged as more permissible than others regardless of human or AI authorship. While this relative ordering remains largely stable, AI-generated transformations are consistently rated as less acceptable than identical human-produced changes. These results reveal a distinction between more permissible and more sensitive alterations, and suggest the need for both designers and AI-assisted visualization tools to incorporate constraints and guardrails that reflect these differences.

} 

\keywords{Cinematic scientific visualization, science communication, visualization aesthetics, generative AI.}

\begin{document}



\maketitle

\section{Introduction} 
Improving the appearance of a visualization can enhance memorability \cite{usefulJunk,borkinMemorable}, learning \cite{aesthetic-neuroscience}, credibility \cite{scivis_aesthetic_grounds_for_trust, vis_aesthetics_influence_trust}, interpretability \cite{effect-aesthetic-interpretation}, and usability \cite{aesthetic-usability, aesthetics-and-usability}. These benefits are especially relevant for communicating scientific results to audiences broader than domain experts. However, transforming a traditional or analytical visualization into one suitable for public communication may involve a wide range of modifications, from relatively superficial changes at the level of presentation (e.g.,~camera position, lighting, or background) to alterations at the data level (e.g.,~denoising, feature omission/addition). Some of these changes may affect how faithfully the visualization represents the underlying data or phenomenon. 

Recent advances in artificial intelligence (AI) introduce new ways to apply transformations (e.g.,~\cite{paraview-mcp, transfer-functions-ai, StyleRF-volvis}). These systems lower the barrier to producing aesthetic modifications, but also introduce new challenges such as limited controllability \cite{how-designers-use-genAI} and potential hallucination \cite{hallucination-survey}.
As AI tools become integrated into visualization workflows, there is a growing need to define the constraints and guardrails for applying such transformations. Rather than imposing assumptions about which AI features should be supported or restricted, we ground this problem in expert judgment by examining which aesthetic alterations are considered acceptable. 
We use \textit{acceptable} to describe judgments about whether a transformation is appropriate for a given context, recognizing that such judgments may reflect a combination of concerns including accuracy, interpretability, and the potential to mislead. This framing also allows us to examine if these judgments shift depending on whether the modifications are applied by a human or an AI system.

We conducted a survey study (N=95) in which visualization researchers, practitioners, and domain scientists assessed fifteen visualization modifications applied by a human designer, and then evaluated the same alterations when generated by an AI system. Participants also assessed the use of AI-enhanced visualizations across communication contexts and rated the importance of factors such as disclosure of AI use and explainability.

Results show that acceptability is driven more by how a transformation affects meaning than the level at which it occurs in the visualization pipeline, with consistent groupings of more and less acceptable modifications. AI-generated alterations are rated less acceptable, and judgments vary by communication context and by factors such as expert validation, disclosure, and understanding the data-to-visual mapping.

Our contributions are: 
(1) An empirical characterization of acceptable and unacceptable aesthetic modifications in scientific visualization; 
(2) A comparison of perceptions of human-created versus AI-generated modifications; 
(3) Implications for AI-assisted visualization tools, emphasizing interpretable, verifiable transformations.

\section{Related Work}

Prior research has examined how visual design and generative AI shape evaluation and interpretation.

\textbf{Aesthetics.} 
Prior work typically evaluates aesthetics at the level of complete visualizations, often using subjective ratings of visual quality and relating these judgments to downstream outcomes such as recall, interpretation, or persuasion (e.g.,~\cite{usefulJunk, borkinMemorable, aesthetic-usability, persuasive-vis}). 
Our work differs by examining elementary modifications (e.g.,~smoothing, lighting, feature edits) in isolation, rather than overall positive or negative aesthetic ratings of complete graphics.

\textbf{Misleading visualizations.} Research on misleading visualizations has identified specific problematic practices, such as deceptive scales \cite{deceptive-vis-techniques, deceptive-impact-vis}, misleading summarization \cite{error-bars-harmful, beyond-graphs}\add{,} and perceptually problematic encodings \cite{rainbow-harmful, graphical-perception}. However, empirical studies typically examine techniques as individual factors, often in the context of statistical charts. As a result, it provides limited insight into how a broader range of transformations---particularly in 3D scientific visualization---compare to one another.

\textbf{Perceptions of generative AI.} Generative AI not only expands how visualizations can be produced \cite{genAI4vis, doom-deliciousness, StyleRF-volvis, genAI4vis-survey}, but also shapes how they may be evaluated: identical outputs are perceived as less credible or appealing when attributed to AI rather than humans \cite{ai-literacy-and-attitudes}, and errors are judged as more unacceptable when attributed AI than to a scientist \cite{Freiling2026}. In addition, AI-based workflows can reduce transparency around how outputs are produced, making transformations harder to parameterize, reproduce, or audit \cite{explainable-genAI, pitfalls-of-genAI}. These factors suggest that both the perceived source of a transformation and the opacity of AI processes can influence how visualization transformations are judged.

\section{Methodology}

We conducted \remove{interviews and }a survey to characterize expert perspectives on the acceptability of aesthetic modifications to 3D scientific visualizations. Our study was determined exempt by the University of Utah's Institutional Review Board (\#IRB\_00199457).

\noindent \remove{ \\ \textbf{3.1 Motivating Interviews} \\ Prior to survey design, we conducted semi-structured interviews with three domain scientists (S1--S3) to scope the problem space. Interviewees were asked general questions about tools and processes for visualizations for communication, including the use of AI. In the second half of the interview, they were shown AI-enhanced versions of their own visualizations for discussion. S1 was enthusiastic about possible tools and the AI-enhanced images, noting that no visualization of their data can be fully accurate as the science is unknown. S2 took a more pragmatic stance in the trade-off between accuracy and clarity, preferring the AI-enhancements only when not for publication. S3 represented a more data-fundamentalist view with regard to what could be `real or true' and was uncomfortable with the black-box behavior of AI. This variation in how visual modifications were interpreted and received motivated our broader survey-based investigation. }

\subsection{Survey Design} \label{sec:survey-design}

We designed an exploratory survey to examine expert perceptions of visual modifications and the use of AI-enhanced imagery in visualization. The survey consisted of three phases, with optional open-ended responses at the end of each section. 

\textbf{(1) Teapot Phase.} Participants were shown fifteen types of alterations applied to an abstract 3D teapot scene (see~\cref{fig:teaser}, top) spanning both data-level and presentation-level changes. \add{Each stimulus was presented as an animated .gif image alternating between the original and modified visualization so that the change could be easily detected. The alteration category was explicitly labeled. Alteration types were presented in random order. The examples were manually created in Houdini by one of the authors as representative instances of each transformation category, and they were iteratively refined through team feedback to ensure recognizable, plausible enhancements within a common scene. The}  abstract scenario was used to reduce fixation on any particular dataset or domain. Although some alteration types can produce semantically or visually similar outcomes, we treated them as distinct because they involve different operations and implications. For example, changing the camera position versus changing the position of the data can result in an identical image, but may lead to different judgments of acceptability. Participants indicated whether each alteration would be acceptable on a six-point Likert scale, first completing the set for a human performing the changes and then again for AI. \add{The same visual stimuli were shown in both conditions; only the stated source of the modification changed.}
An even number of response options was used to encourage participants to indicate a directional leaning.

\textbf{(2) Context Phase.} Participants were next shown pairs of exploratory and AI-enhanced visualizations (\cref{fig:teaser}, bottom) and asked to assess their appropriateness across communication contexts (e.g.,~publications, outreach) and the importance of factors such as expert validation and disclosure of AI use. These responses were collected on a 5-point Likert scale, 
using an odd number of options to allow a middle response for uncertain evaluations of appropriateness and importance.

\begin{figure}
\centering
  \includegraphics[width=\linewidth]{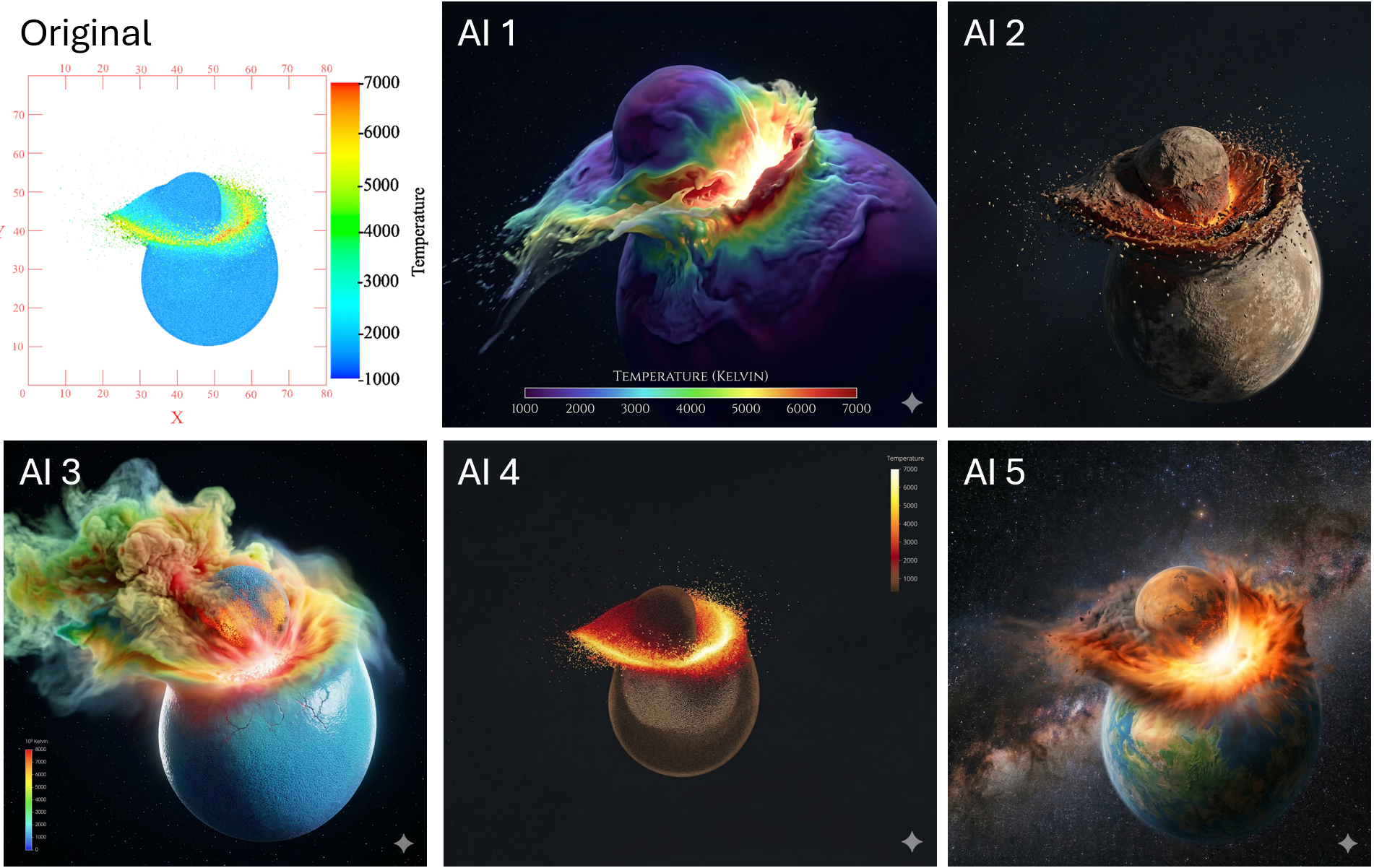}
  \caption{\remove{AI stimuli used in the survey. }The original \add{planetary collision} visualization \add{and} 
  \remove{is shown alongside }five AI-generated variations that introduce aesthetic modifications \add{used in the survey}. Participants were asked to describe aspects of these images that would make them comfortable or uncomfortable if presented as visualizations. \small{\textit{AI images generated with Nano Banana Pro.}}}
  \label{fig:planets}
\end{figure}

\textbf{(3) Case Study Phase.} Finally, the survey presented a domain-specific example: an analytical visualization of a planetary collision alongside five AI-generated variations (\cref{fig:planets}), with modifications ranging from minimal to highly stylized. Participants responded to an open-ended prompt asking what would make them comfortable or uncomfortable calling these images ``data visualizations.'' This prompt was intended to elicit discussion of judgments and the factors leading to them.

\subsection{Participants}

Participants were recruited via email lists and social media, resulting in 95 responses. 
Participants identified as visualization researchers (27), visualization practitioners (24), scientists who create visualizations (24), and scientists who use visualizations created by others (20). Participants spanned a range of scientific domains, with the strongest representation from astronomy (56), earth/geoscience (48), and engineering (35) and with most working across multiple domains (72). The sample spanned ages 18--70 and was predominantly male (68 male, 18 female, 1 non-binary, 8 prefer not to answer), highly educated (48 doctorate, 25 master's, 19 bachelor's, 2 some college), and White (57) followed by Asian (18). Participants reported regular engagement with visualization (61 at least weekly, including 41 daily) and substantial experience (70 with more than 5 years), including experience creating visualizations for the general public (39 occasionally, 38 frequently or as a primary part of their work). Most participants had experience with generative AI for image generation (80). \add{We relied on participant self-reporting and did not test or verify expertise or experience.}

\subsection{Analysis}

Quantitative responses were summarized using descriptive statistics.
To examine consistency and variation in judgments, we compared responses across participant subgroups (e.g.,~role, domain, experience level) and alteration source (human vs.~AI). We analyzed relative ordering and agreement in acceptability judgments to identify shared patterns and systematic differences.

Open-ended responses were analyzed in two stages. We deductively coded the responses by the 15 alteration types introduced in the survey and report these in~\cref{sec:deductive-coding}. Three coders independently labeled each response for positive, negative, or other mentions of these categories. \add{Inter-coder reliability was moderate, with Krippendorff's $\alpha$ = 0.69.} Disagreements were resolved through discussion to reach a consensus.

Second, we conducted two rounds of inductive coding to identify cross-cutting concerns and evaluation criteria which were not captured by the predefined categories, reported in~\cref{sec:inductive-coding}. One author generated an initial set of codes, which were refined through discussion with the full coding team.

\subsection{Limitations}

This study was exploratory, and the fifteen transformation types were selected by the authors rather than derived from a formal taxonomy. These categories have overlapping, context-dependent meanings (see~\cref{sec:survey-design}), which limits the precision of the distinctions between them. As a result, the reported rankings should be understood as relative judgments within this set rather than a comprehensive or strictly defined ordering of all possible transformations. The set is also incomplete: participants referenced additional transformations in the open-ended responses---scaling and slicing---which were not included in our survey questions. \add{Participants may have also reasonably interpreted terms such as smoothing, enhancement, or feature addition differently depending on their disciplinary background and visualization experience.} Finally, the use of abstract stimuli was intended to reduce domain-specific bias but may limit ecological validity, and \add{because each transformation category was represented by a single example, responses may partially reflect the specific rendering chosen}.

\section{Results}

We present quantitative patterns in acceptability, then use qualitative responses to explain how these judgments are reasoned about, and examine how acceptability varies by context.
Data can be explored at {\small \url{https://kalinalinkalina.github.io/beautify-vis-results}}. 

\subsection{Acceptable and Unacceptable Alterations}

\begin{figure}
\centering
\includegraphics[width=\linewidth]{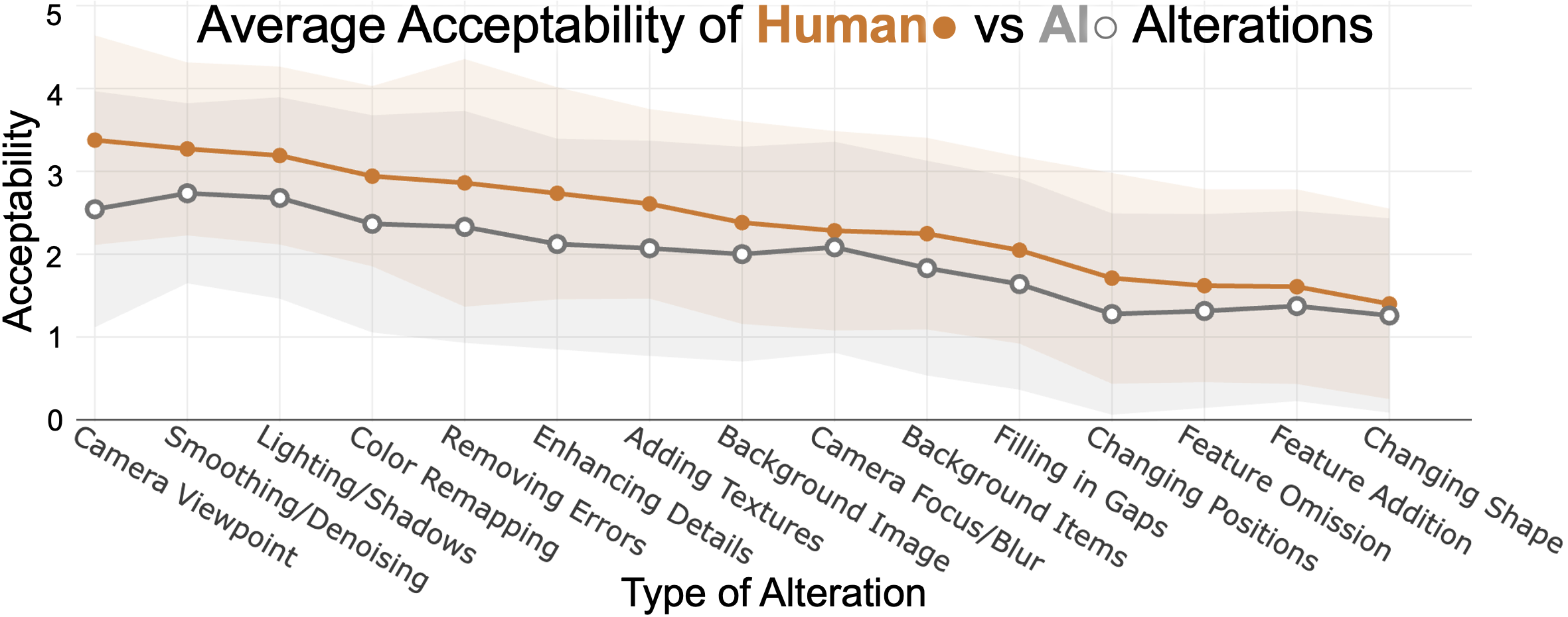}
\caption{Mean acceptability (±1 SD) for human and AI alterations, ordered by human ratings. Human alterations are rated higher across all types. The gap narrows for low-acceptability changes.}
\label{fig:line-chart}
\end{figure}

We begin by examining patterns across transformation types, based on responses in the teapot phase. Results are summarized in~\cref{fig:line-chart}.

\textbf{Acceptability varies by transformation type.} We use ratings of human-generated alterations as the baseline for acceptability. Camera position, smoothing, and lighting received the highest ratings (\textit{M} = 3.38, 3.27, 3.19, respectively), while changing position, feature omission, feature addition, and changing shape received the lowest ratings (\textit{M} = 1.71, 1.62, 1.61, 1.40). The remaining alterations fell between these extremes (\textit{M} = 2.05--2.94).

\textbf{AI lowers acceptability but preserves ordering.} AI-generated alterations rated lower than human-made across all categories, and the magnitude of this difference varies by alteration type.  
The relative ordering of alteration types remains largely stable across human and AI alterations: the same groups of transformations appear at the top and bottom of the ranking, with minor reordering within these groups. Least acceptable changes were rated similarly low for both humans and AI, whereas the most acceptable changes were rated higher for humans than for AI. Qualitative responses suggest this reflects reduced confidence in AI: humans were assumed to have intent, accountability, and domain understanding, while AI was assumed to have opacity, hallucination, and a lack of context.

\textbf{Acceptability depends on interpretation, not where a change occurs.} The distinction between acceptable and unacceptable transformations is not simply bifurcated by \textit{presentation-level} changes (e.g.,~camera, lighting, color) versus \textit{data-level} changes (e.g.,~smoothing, feature addition/omission). While presentation-level changes are often rated as more acceptable than data-level changes, this pattern breaks down in a key case: smoothing, an alteration to the data, is among the most acceptable transformations, whereas camera blur, which does not alter the data, is rated substantially lower. This smoothing/blur paradox \remove{indicates}\add{suggests} that acceptability is not determined by where a transformation occurs in the pipeline, but by how it affects the interpretation of the data.

\subsection{Transformation-Specific Judgments} \label{sec:deductive-coding}

In deductive coding of open-ended responses, participants described lower-ranked transformations, particularly feature addition and shape changes, as altering meaning, noting that such changes \textit{``generate new shapes, features, or falsify data''} and that \textit{``anything that misleads or renders an incomplete/incorrect shape of the data is rarely acceptable.''} In contrast, positive comments were sparse\remove{primarily associated with higher- and mid-ranked transformations}, typically describing changes that enhance clarity or perception without affecting interpretation.

Participants discussed whether a transformation affected interpretation, with even feature addition considered acceptable \textit{``when the feature added has no bearing on the interpretation of the primary dataset.''} This pattern was summarized by one participant: \textit{``some changes are always fine (ie don't adjust the data or potential change interpretation), while others are contextual (is it going to change how the data is understood or not in the context).''} Comments on color changes highlight this dependence: while rated as acceptable in the abstract teapot phase of the survey, they prompted strong negative reactions in the case study (\cref{fig:planets}), when encodings were perceived as unreliable (e.g.,~\textit{``I would also be very very uncomfortable if the visualization color transfer function is hallucianted by AI''}), revealing a gap between acceptability in principle and in practice given current AI limitations.

\subsection{Reasoning About Acceptability} \label{sec:inductive-coding}

Beyond specific transformations, participants described broader principles guiding acceptability. Truth and accuracy were common concerns, with one writing, \textit{``any modification that significantly distorts the reality / truth of the data is unacceptable.''} Several participants focused on being \textit{``truthful to the data''} in particular, and some participants viewed some aesthetic changes or AI-enhanced images as altering the data. 

Participants saw purpose in aesthetic changes, with one participant expressing, 
\textit{``you can work the visuals to make it more aesthetic or emphasize a point in the data to make a particular message.''} Many pointed out that the acceptability was highly tied to task and purpose. While several participants acknowledged that communicating effectively to general audiences may require selective abstraction or emphasis; only one explicitly stated that being \textit{``convincing or clear is more important than strict accuracy.''} Improving clarity or comprehension was a common reason to make aesthetic changes. These included emphasizing important features, reducing abstractions, and providing context for the audience. However, several participants cautioned that even in these cases, the audience might not interpret the finished product as expected.

Participants noted that if the result was the same, it did not matter if the change was by a human or AI, especially with human oversight and validation. However, participants also expressed distrust of AI to \textit{``get it right,''} suggesting that it can introduce errors and lacks
context to produce acceptable changes. Trust continued to be a concern, with one participant writing, \textit{``Even if validated by an expert, it reduces public trust.''} Others expressed trust would increase if they knew more about the AI and how it had been validated.

Several participants viewed color bars, axes, and legends as signifiers of being a data visualization with a few suggesting the lack of any of these made the images \textit{``something other than ``data visualizations.''''} Participants also found several of the AI-generated images to be an \textit{``artist's impression''} or illustration. One respondent warned that whether AI was used or not, \textit{``photorealism can easily get in the way of accurately representing data and should then not be used''} and another cautioned against \textit{``natural''} depictions of extreme events because it \textit{``implies details that are not known.''}

\subsection{Contextual Factors in Acceptability Judgments}

We return to quantitative results to examine how acceptability varies across participant groups and contexts.

\textbf{Effects of participant characteristics.} Acceptability ratings varied across participant subgroups. 
Some patterns followed intuitive trends: acceptance increased strongly with willingness to use AI visualization enhancement tools and, to a lesser extent, among less experienced and younger participants. 
A notable pattern emerged by professional role: responses were similar for human-made alterations but diverged for AI-generated ones, with visualization researchers more accepting than practitioners (\textit{M}=2.10 vs 1.85 on a scale 0--5, and scientists who use visualizations more accepting than those who create them (\textit{M}=2.31 vs 1.66). 

\textbf{Acceptability across contexts.} Ratings were moderate for informal contexts such as social media posts, public talks, and press releases (\textit{M}=2.21, 2.11, 1.91 on a scale 0--4); low for conference presentations and grant proposals (\textit{M}=1.33, 1.33); and extremely low for research papers (\textit{M}=0.96).

\textbf{Important conditions for acceptability.} Disclosure of AI use, expert validation, clarity about what changes were made and why, and understanding the data-to-visual mapping were rated highly important (\textit{M}=3.44, 3.43, 3.37, 3.35 on a scale 0--4). The ability to revise or iterate was also valued (\textit{M}=3.19). Providing detailed information about the AI process was less important (\textit{M}=2.42).

\subsection{Willingness to Use AI Visualization Tool}

At the end of the survey, participants were asked whether they would use an AI visualization enhancement tool, if one were available to them. Despite moderate acceptability ratings earlier in the survey, responses indicated openness to use: 51 responded ``Maybe'', 27 ``Yes'',  and 13 ``No.'' In the qualitative responses, two participants mentioned already using AI in this way.

\section{Discussion \add{and Conclusion}}

We interpret these findings and discuss implications for visualization design and AI-assisted tools.

\textbf{Considerations for acceptability.} Designers making aesthetic or narrative alterations to a visualization should evaluate transformations along three dimensions 
(1) \textit{Effect on data meaning}: designers can be more flexible when applying transformations that clarify or preserve structure and are broadly accepted, but lower-ranked changes should be used sparingly and require stronger justification or domain expert validation as they may distort, obscure, or fabricate features.
(2) \textit{Interpretability}: ensure that changes can be understood and justified, as participants prioritized knowing what was changed and why over details of the underlying process.
(3) \textit{Context of use}: consider the \remove{risk of misinterpretation in}\add{norms and expectations of} the target setting, as transformations acceptable in one context may be viewed as \remove{misleading}\add{inappropriate} in another; apply stricter constraints in analytical settings than in communicative ones.

\textbf{What you call it changes what is acceptable.} In addition to the context of use, acceptability also depends on how an image is framed. Participants applied stricter standards if they perceived an image to be a ``visualization'', and greater tolerance for alterations in ``illustration''. Although the boundary between these categories is ambiguous in practice \cite{picture-perfect-scicomm, illustration-techniques-4-vis, illustrative-visualization}, experts treated it as meaningful when evaluating acceptability. Transformations that push a visualization toward illustration may still be acceptable in communicative contexts, but they require clear signaling to avoid misinterpretation.

\textbf{Early signals of changing norms.} 
Acceptability remains low even in the most permissive contexts (e.g.,~ social media, \textit{M}=2.21), yet reported attitudes and behavior suggest emerging adoption: only 14\% of participants opposed AI enhancement tools, and two are already using them in practice. This gap between stated acceptability and use mirrors patterns in other domains where image manipulation and beauty filters are widely criticized yet still widely adopted because they produce desirable outcomes such as increased engagement \cite{filtered-reality, beauty-filters-good}. In photography, practices such as retouching, compositing, and filtering have expanded what is considered acceptable manipulation, alongside evolving standards around disclosure and authenticity \cite{can-we-trust-photographs, effectiveness-manipulation-disclosure}. Similarly, beauty filters and social media editing tools normalize continuous, low-effort modification of visual appearance, shifting expectations of realism and influencing how visual content is interpreted \cite{filtered-reality, beauty-filters-good, illegally-beautiful-manipulation-disclosure}. Visualization may be at an early stage of a similar transition, in which \remove{standards}\add{attitudes toward AI-assisted visualization enhancement} are likely to evolve in response to new tools. Our findings on AI acceptability reflect current norms, which are likely to evolve, whereas judgments of human-generated alterations are more likely stable.

\textbf{Implications for design with aesthetic modifications.} 
We identify implications for both visualization design and AI-assisted tools, with future work needed to operationalize them: 

\noindent \textit{Support validation and accountability}, as expert verification of transformations is critical, particularly in higher-stakes contexts; 

\noindent \textit{Constrain high-risk transformations}: operations perceived as altering structure or introducing new features should require stronger controls, explicit signaling, or be restricted by default; 

\noindent \textit{Ensure transformations are legible and reversible}, as acceptable transformations are those that can be understood, justified, and if necessary, undone. 

And for AI tools specifically: 

\noindent \textit{Prioritize reducing the trust penalty}: given that transformation rankings are largely stable, increasing confidence in AI outputs may have greater impact than focusing on individual alterations; 

\noindent \textit{Prioritize mapping over model transparency}, as participants valued understanding what was changed and how it affects the representation over details of the underlying model. 

\acknowledgments{
We thank Francesca Samsel, Arleth Salinas, Chris Johnson, Sebastian Frith, Moose, and Jo. 
This project was funded in part by NSF-2324465, the Intel OneAPI CoE, and the Intel Graphics and Visualization Institutes of XeLLENCE.
}

\bibliographystyle{abbrv-doi}

\bibliography{the-bibliography}

\end{document}